\documentstyle[preprint,aps]{revtex}

\begin{document}
\draft

\title{Internal Spatial Oscillations in a Single Trapped
Bose--Einstein Condensate}
\author{Alexander V. Zhukov}
\address{ Department of Theoretical Physics, Belgorod State
University, 12 Studencheskaya str.,\\ 308007 Belgorod, Russian
Federation}

\date{\today}
\maketitle

\begin{abstract}
I predict the existence of internal spatial currents in a {\it
single} macroscopic quantum system, namely in trapped dilute-gas
at sufficiently low temperatures, when a Bose-Einstein
condensation occurs. The spatial profiles of the wavefunctions of
low-lying states in such a system are different due to the
inhomogeneity, caused by an asymmetry of external trapping
potential. This is the reason for appearing of Josephson--like
oscillations between atomic subsystems in different states
including the ground state as well. Using a simple model for the
wavefunctions of three low-lying states we demonstrate how
essential this effect can be. The possible applications of the
predicted effect are briefly discussed. Particularly, this effect
opens the possibility to identify experimentally the low lying
excited states of a system.
\end{abstract}

The recent observations of Bose-Einstein condensation (BEC) in
trapped atomic gases \cite{1} have renewed interest in bosonic
many-particle systems at low temperatures. The strong
inhomogeneity of trapped gases in real experiments makes a system
to be even more interesting.

\par
Of course, the most spectacular consequences of Bose-Einstein
condensation is superfluidity and high phase coherence \cite{2}.
The latter can lead to quite a number of interference phenomena,
which were not being observed earlier in macroscopic systems in
such a direct way. One of such interference effects was observed
by the MIT group \cite{3}. They used a laser beam to separate an
atomic cloud into two parts. After switching off the confining
potential and the laser, the authors of \cite{3} observed clean
interference patterns in the overlapping region.

\par
The authors of \cite{4} showed a possibility of another
interesting manifestation of the phase coherence, namely they
predicted the existence of Josephson-type effects in a
double--well external potential. In this case a difference of
chemical potentials ,say $\mu_1$ and $\mu_2$, in two condensates
leads to appearance of the Josephson current $I\propto \sin
[(\mu_1-\mu_2)/\hbar t]$. Analytic calculations of the spatial
current distribution are rather complicated because of necessity
to account the boundary conditions of condensates order parameters
\cite{2}, \cite{4}. Nevertheless, the Josephson oscillations are
already observed in experiment \cite{5} and are extensively
studied by various groups \cite{6} both theoretically and
experimentally.

\par
Here we propose a simple model theory for internal spatial
oscillations, which occur in a {\it single} condensate. The origin
of this effect is a difference of spatial profiles of
wavefunctions of the ground state and neighboring excited states.
So, the well--known quantum interference between two quantum
states can lead to the observable spatial process in the systems
under consideration.

\par
It was realized long ago, that the ground state properties of
trapped Bose-gas can be well described by the Gross--Pitaevskii
equation \cite{7}
 \begin{equation}
 i\hbar\frac{\partial\Phi ({\bf r},t)}{\partial\hbar}=
 \left\{
 -\frac{\hbar^2}{2m}\Delta +U_{\rm{trap}}({\bf r})+
 g|\Phi ({\bf r},t)|^2
 \right\}\Phi ({\bf r},t),
 \label{1}
 \end{equation}
where $\Phi ({\bf r},t)=\langle\hat{\psi}({\bf r},t)\rangle$ is
the condensate wavefunction, $U_{\rm{trap}}({\bf r})$ is the
trapping potential,
 \[
 g=\frac{4\pi\hbar^2a}{m}
 \]
is the zero-range interaction constant, $a$ is the $s$-wave
scattering length. To obtain the ground state properties, one can
write the condensate wavefunction as $\Phi ({\bf r},t)=
\varphi({\bf r})\exp(-i\mu t/\hbar)$. In fact, this mean we
assumed a condensate to be in the stationary state. In this case
equation (\ref{1}) takes the stationary form
 \begin{equation}
 \mu\varphi ({\bf r})=
 \left\{ -\frac{\hbar^2}{2m}\Delta + U_{\rm{trap}}({\bf r})+
 g|\varphi ({\bf r})|^2\right\}\varphi
 ({\bf r}). \label{2}
 \end{equation}
This equation explains the sense of the chemical potential $\mu$
as an energy of the stationary level. This can be used further for
determining of $\varphi ({\bf r})$ and $\mu$.

\par
In this Letter we want to study the ground state of trapped bosons
in an asymmetric harmonic potential. For our aims we must allow
for (at least) two single-particle states. Keeping in mind the
illustrational purposes we shall consider the simplest case of
stationary states, where the single--particle wavefunctions have
the oscillator--like form \cite{8}
 \begin{equation}
 \Phi_j({\bf r},t)=\varphi_j({\bf r})
 \exp\left(-i\frac{\mu_jt}{\hbar}\right),
 \label{3}
 \end{equation}
where
 \begin{eqnarray}
 \varphi_0({\bf r})=\frac{1}{\pi^{3/4}b^{3/2}}
 \exp\left(-\frac{r^2}{2b^2}\right),\label{4}\\ \varphi_1({\bf r})=
 \frac{\sqrt{2}}{\pi^{3/4}b^{5/2}}z
 \exp\left(-\frac{r^2}{2b^2}\right),\label{5}
 \end{eqnarray}
$b$ is a variational parameter. These have the same shape as the
ground state and the first excited state with a zero projection of
angular momentum. More generally, we could take $\varphi_1 ({\bf
r})$ as a linear combination of the states with different angular
momentum projections $m_l=0,\pm 1$, but this has no influence on
the qualitative picture. Furthermore, below we will operate in
fact with the one dimension, say $z$. This is particularly
acquitted in view of spatial anisotropy of trapping potential in
majority of real experiments \cite{9}. The resulting wavefunction
of our simple model system can be presented as a linear
combination:
 \begin{equation}
 \Phi^{(01)}({\bf r},t)=\varphi_0({\bf r})
 \exp\left(-i\frac{\mu_0}{\hbar}t\right) +
 \varphi_1({\bf r})\exp\left(-i\frac{\mu_1}{\hbar}t\right).
 \label{6}
 \end{equation}
The current density is then given by the well--known relation
\cite{10}
 \begin{equation}
 {\bf I}^{(01)}=\frac{i\hbar}{2m}\left\{ \Phi^{(01)}({\bf r},t)
 \nabla\Phi^{(01)*}({\bf r},t)- \Phi^{(01)*}({\bf r},t)
 \nabla\Phi^{(01)}({\bf r},t)\right\}.
 \label{7}
 \end{equation}
Substituting eq. (\ref{6}) into eq. (\ref{7}), we come to
 \begin{equation}
 {\bf I}^{(01)}=-\frac{\hbar}{m}\varphi_0({\bf r})\varphi_1({\bf r})
 \frac{{\bf e}_z}{z}\sin\left(\frac{\mu_1-\mu_0}{\hbar}t\right),
 \label{8}
 \end{equation}
where ${\bf e}_z$ is the unit vector along $z$--direction.
Equation (\ref{6}) looks very similar to the corresponding
expression in \cite{4} for the Josephson current between two
condensates, but have rather different sense. This simple example
shows that there is an oscillating current which results in
oscillations of particle number in each state. For example the
temporal variation of the number of ground state particles due to
a Josephson-type exchange with the first excited state is given by
the formula
 \begin{equation}
 \delta n_0^{(1)}(z,t)
 =\frac{2^{3/2}\hbar^2z}{\pi^{3/2}mb^6(\mu_1-\mu_0)}
 \exp\left(-\frac{z^2}{b^2}\right)\cos
 \left(\frac{\mu_1-\mu_0}{\hbar}t\right). \label{9}
 \end{equation}
Typical dependence of $\delta n_0^{(1)}(z,t)$ on $z$ and $t$ is
presented in Fig. \ref{f1}.

\par
Significance of the discussed effect is especially clearly
manifests if we account the second excited state as well. Taking
the simple oscillator-type wavefunction \cite{10} for this second
excited state, we easily get the following formula
 \begin{equation}
 \delta n_0^{(2)}(z,t)
 =-\frac{2^{5/2}\hbar^2z(1-z^2/b^2)}{\pi^{3/2}mb^6(\mu_2-\mu_0)}
 \exp\left(-\frac{z^2}{b^2}\right)\cos
 \left(\frac{\mu_2-\mu_0}{\hbar}t\right), \label{10}
 \end{equation}
where $\mu$ stands for the chemical potential of the second
excited level. Fig. \ref{f2} shows the dependence of a total
variation of the particle number in the ground state $\delta n_0
(z,t)=\delta n_0^{(1)}(z,t)+\delta n_0^{(2)}(z,t)$ on time (in
units of $\hbar/(\mu_1-\mu_0)$), calculated from expressions
(\ref{9}) and (\ref{10}). The figure clearly illustrates possible
applications of the predicted effect. Namely, if we measure the
oscillations of the number of particles in the ground state (at
any convenient coordinate), then we get a possibility to recover
the low lying excited stationary states ({\it i.e.} to find their
chemical potentials, spreading, spatial shape etc.). This becomes
even more tempting in view of a difference in spatial dependences
of $\delta n_0^{(1)}$ and $\delta n_0^{(2)}$, which is shown on
Fig. \ref{f3}. Thus, all excited states become well
distinguishable by measuring the amplitude and period of
oscillations. Fig. \ref{f4} represents $\delta n_0$ as a function
of both $z$ and $t$. So, the character of oscillations is rather
different for each mode.

\par
In conclusion, using a simple model for describing the ground and
low-lying excited states of the trapped Bose-gas, we showed the
principal possibility of the spatial oscillations within a system
due to the quantum interference between states with different
spatial profiles. These oscillations demonstrate rather different
behaviour for different energy levels of the system. This fact
provides a tempting possibility to determine the spectrum of
stationary states in the system.

\par
In reality the wavefunctions are of course different from those
used above due to an interaction. Furthermore, we even do not
certainly know whether or not the stationary excited states with
well defined chemical potentials exist. The measuring of the
considered internal oscillations just can answer this question.

\begin{figure}
\caption{The dependence of $\delta n_0^{(1)}$ on $z$ and t (here
$\tau=(\mu_1-\mu_0)t/\hbar$). We used the constants $b= 5\cdot
10^{-3}$ cm, $\mu_0=830\hbar$ erg, $\mu_1=1.3\cdot 10^3\hbar$ erg.
Note, the real amplitude of oscillations may be, of course, not so
large. The parameters of wavefunctions (\ref{3}) are taken rather
approximately to draw the general picture.} \label{f1}
\end{figure}

\begin{figure}
\caption{Temporal dependence of $\delta n_0=\delta
n_0^{(1)}+\delta n_0^{(2)}$ at $z=b/2$
($\tau=(\mu_1-\mu_0)t/\hbar$). The constants are $b= 5\cdot
10^{-3}$ cm, $\mu_0=830\hbar$ erg, $\mu_1=1.3\cdot 10^3\hbar$ erg,
$\mu_2=3.6\cdot 10^3\hbar$ erg.} \label{f2}
\end{figure}

\begin{figure}
\caption{Spatial dependence of $\delta n_0^{(1)}$ and $\delta
n_0^{(2)}$ at $\tau=(\mu_1-\mu_0)t/\hbar=6$. All constants are the
same as in Figs. \ref{f1} and \ref{f2}.} \label{f3}
\end{figure}

\begin{figure}
\caption{The dependence of $\delta n_0=\delta n_0^{(1)}+\delta
n_0^{(2)}$  on $z$ and $t$. All constants are the same as in Figs.
\ref{f1} and \ref{f2}.} \label{f4}
\end{figure}


\begin{references}
\bibitem{1}M.H. Anderson, J.R. Ensher, M.R. Matthews, C.E. Wieman,
           and E.A. Cornell, Science {\bf 269}, 198 (1995);
           K.B. Davis, M.-O. Mewes, M.R. Andrews, N.J. van Druten,
           D.S. Durfee, D.M. Kurn, W. Ketterle, Phys. Rev. Lett.
           {\bf 75}, 3969 (1995); C.C. Bradley, C.A. Sackett,
           J.J. Tollet, R.G. Hulet, Phys. Rev. Lett. {\bf 75}, 1687 (1995);
           C.C. Bradley, C.A. Sackett, R.G. Hulet, Phys. Rev. Lett.
           {\bf 78}, 985 (1997)

\bibitem{2}F. Dalfovo, S. Giorgini, L.P. Pitaevskii, and S. Stringari,
           Rev. Mod. Phys. {\bf71}, 463 (1999); L.P. Pitaevskii,
           Usp. Fiz. Nauk {\bf 168}, 641 (1998)

\bibitem{3}M.R. Andrews, C.G. Townsend, H.-J.~Miesner, D.S.~Durfee,
           D.M.~Kurn, and W.~Ketterle, Science {\bf 275}, 637 (1997)

\bibitem{4}F. Dalfovo, L. Pitaevskii, and S. Stringari,
           Phys. Rev. A {\bf 54}, 4213 (1996); cond-mat/9604069

\bibitem{5}J.I. Cirac {\it et al.}, Phys. Rev. A {\bf 57}, 1208
           (1998); S. Raghavan, A. Smerzi, and V.M. Kenkre, {\it
           ibid}. {\bf 60}, R1787 (1999); J. Williams {\it et
           al}., {\it ibid}. {\bf 59}, R31 (1999); A. Sinatra and
           Y. Castin, Eur. Phys. J. D {\bf 8}, 319 (2000)

\bibitem{6}A.J. Leggett, Rev. Mod. Phys., to appear; D. Jaksch {\it et
           al}, cond-mat/0101057

\bibitem{7}E.P. Gross, Nuovo Cimento {\bf 20}, 454 (1961); L.P.
           Pitaevskii, Zh. Eksp. Teor. Fiz. {\bf 40}, 646 (1961)
           [Sov. Phys. JETP {\bf 13}, 451 (1961)]

\bibitem{8}{\O}. Elgar{\o}y and C.J. Pethick,
           Phys. Rev. A {\bf 59}, 1711 (1999)

\bibitem{9}E.A. Cornell, J.R. Ensher, and C.E. Wieman, cond-mat/9903109

\bibitem{10}L.D. Landau and E.M. Lifshitz, {\it Quantum Mechanics}
            (Pergamon Press, Oxford, 1965)

\end{references}
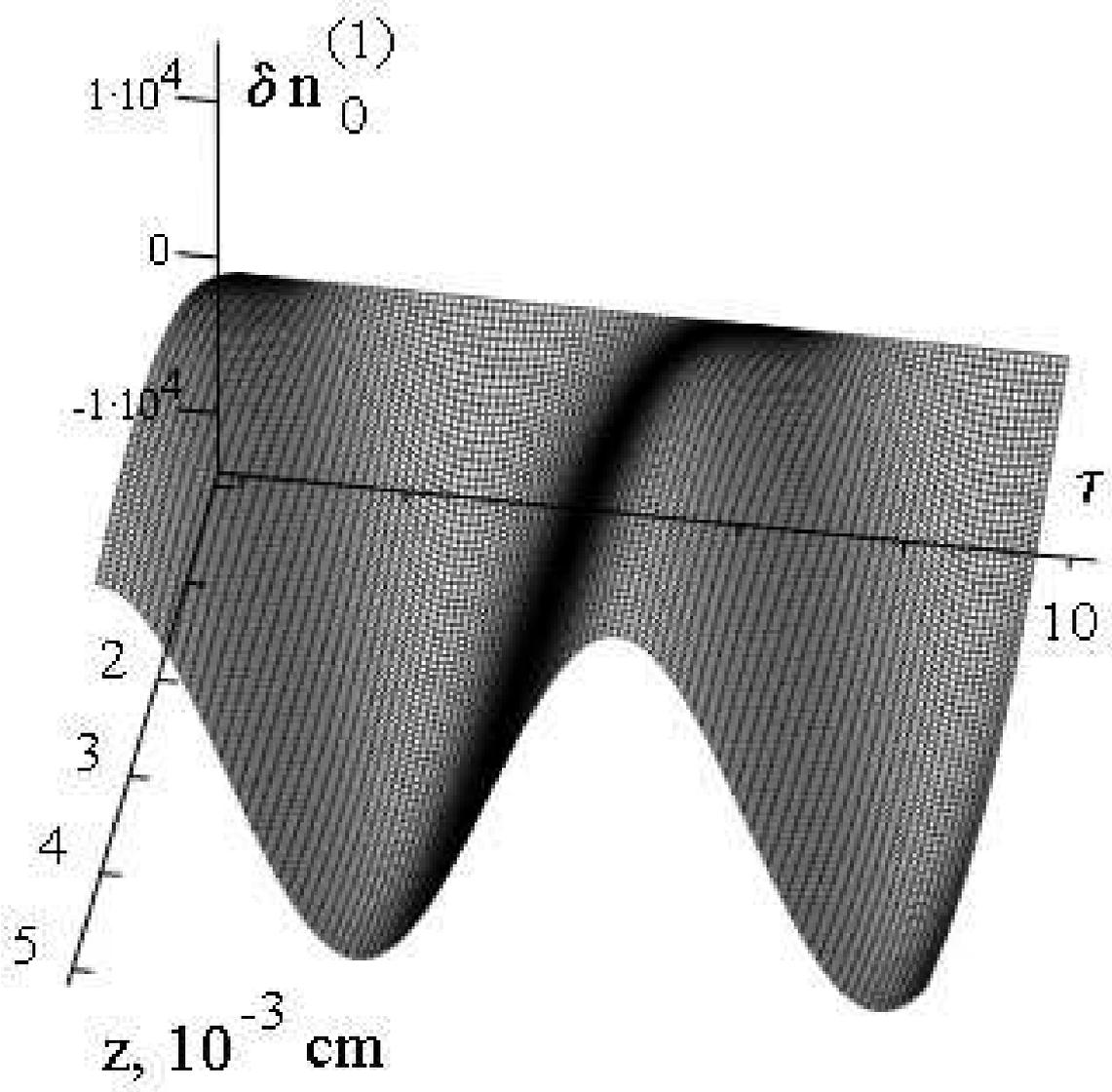
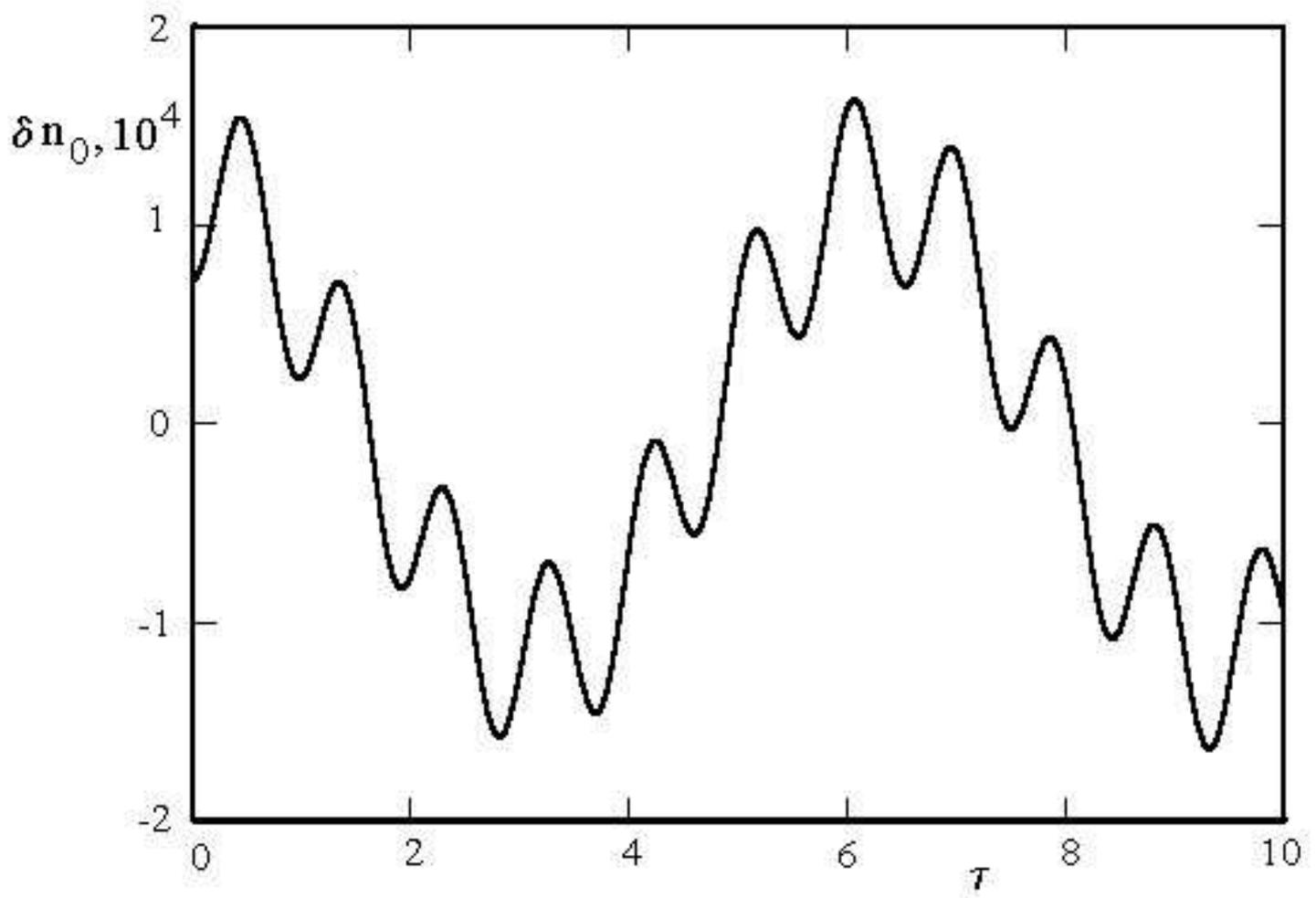
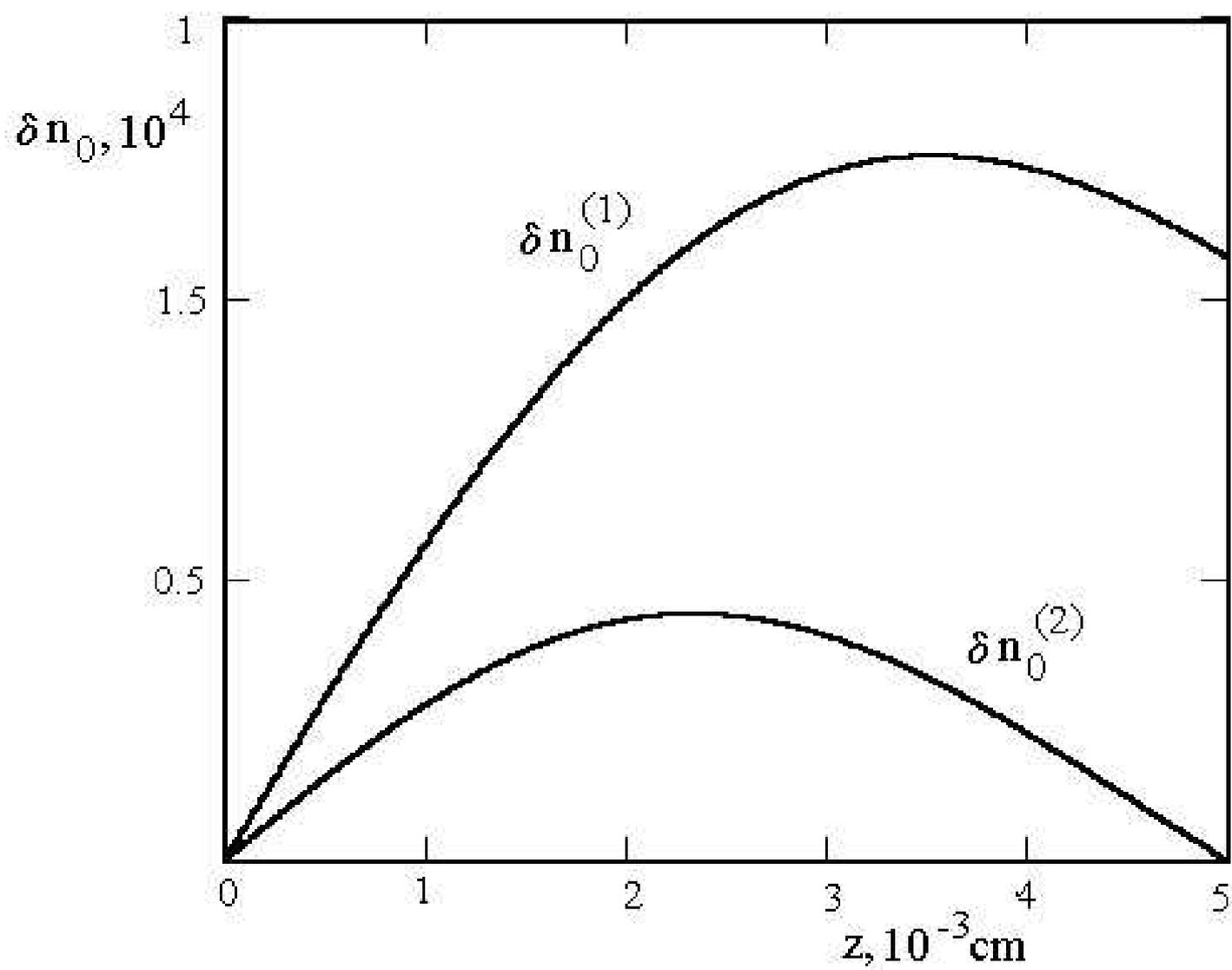
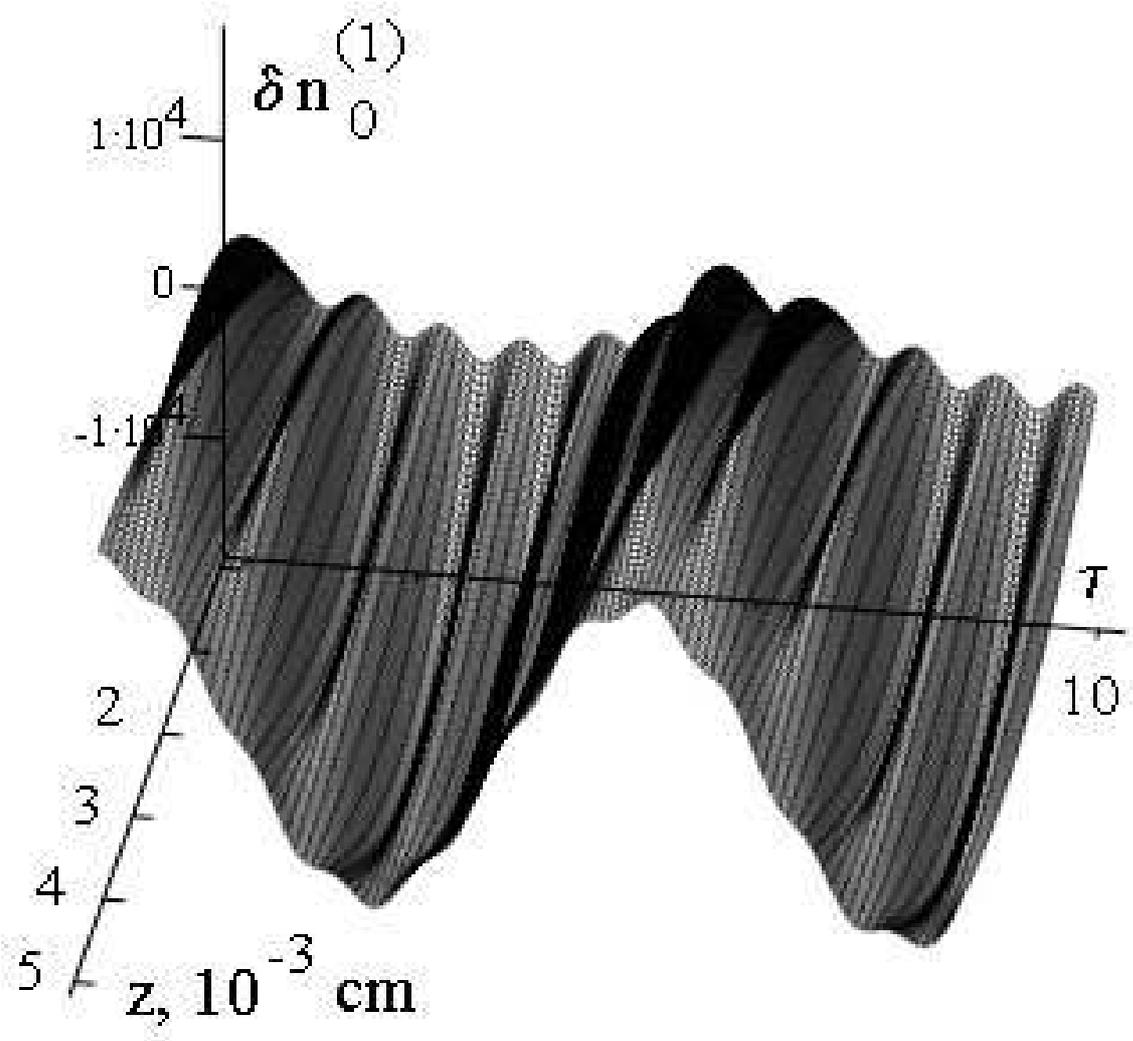
\end{document}